\documentclass[11pt,draftcls,onecolumn]{IEEEtran}
\usepackage[utf8]{inputenc}
\newcommand{\bZ}{{\bf Z}}
\newcommand{\bX}{{\bf X}}
\newcommand{\bT}{{\bf T}}
\newcommand{\bA}{{\bf A}}
\newcommand{\bB}{{\bf B}}
\newcommand{\bC}{{\bf C}}

\newcommand{\bF}{{\bf F}}
\newcommand{\bG}{{\bf G}}

\newcommand{\bH}{{\bf H}}
\newcommand{\bL}{{\bf L}}
\newcommand{\bbl}{{\bf l}}
\newcommand{\bU}{{\bf U}}
\newcommand{\bbu}{{\bf u}}

\newcommand{\bS}{{\bf S}}
\newcommand{\bI}{{\bf I}}
\newcommand{\bY}{{\bf Y}}

\newcommand{\bff}{{\bf f}}

\newcommand{\bc}{{\bf c}}
\newcommand{\by}{{\bf y}}
\newcommand{\bx}{{\bf x}}
\newcommand{\bw}{{\bf w}}
\newcommand{\bv}{{\bf v}}
\newcommand{\bz}{{\bf z}}

\newcommand{\bs}{{\bf s}}
\newcommand{\tT}{{\cal T}}
\newcommand{\tF}{{\cal F}}

\usepackage{stmaryrd}
\usepackage{amssymb}
\usepackage{hyperref}

\begin{document}


%
%
%
\begin{center}

{\huge Characterization of Decomposition of Matrix Multiplication Tensors}

{\large Petr Tichavsk\'y}

{\em The Czech Academy of Sciences, Institute of Information Theory and
Automation,\\ Prague 182 08, Czech Republic}
\end{center}
\vspace*{5mm}

\begin{abstract}
In this paper, the canonical polyadic (CP) decomposition of tensors that corresponds to matrix multiplications is studied.
Finding the rank of these tensors and computing the decompositions is a fundamental problem of algebraic complexity theory. In this paper, we
characterize existing decompositions (found by any algorithm) by certain vectors called signature, and transform them in another decomposition which can be more suitable in practical algorithms. In particular, we present a novel decomposition
of the tensor multiplication of matrices of the size
$3\times 3$ with $3\times 6$ with rank 40.
\end{abstract}

%
%

\section{Introduction}

Computing the minimum number of scalar multiplications needed to calculate the product of two matrices is a fundamental problem of algebraic complexity theory \cite{Landsberg}.
Since the pioneering work of Strassen \cite{Strassen} we know that the complexity of computing the product of two matrices of the size $N\times N$ is at most
$O(N^{2.81})$. This asymptotic bound was improved several times, most recently by Coppersmith and Winograd \cite{Coppersmith}, Williams \cite{Williams}, Le Gall \cite{LeGall}, and Alman and Williams \cite{laser}. The current best asymptotic complexity is $O(N^{2.37286})$.

This paper is devoted to complexity of matrix multiplications (MM) in general. The complexity is expressed as rank of certain tensors, called matrix multiplication tensors or, equivalently, as a solution to the so-called Brent equation \cite{Brent}. We propose a novel matrix formulation of the equation.

A specific problem is the complexity of small matrix multiplication or rank of the corresponding tensor. In the case of matrices of the size $2\times 2$, the tensor has the size $4\times  4\times 4$ and it was proved that the rank is 7 \cite{Winograd}.
However, for matrices of the size $3\times 3$, we only know the bounds for this rank \cite{Conner}. The lower bound is 19 and the upper bound is 23.
A lot of effort has been exerted to improve the algorithm of Laderman \cite{Laderman} (decomposition with rank 23), there are many algorithms with the same complexity \cite{small,Courtois,Oh,Heule,Berger}, but the
proof that a decomposition to a rank smaller than 23 is impossible is not known yet.
For more introductory texts to the subject see, e.g.,  \cite{Bini,Laderman,Blaser,Dumas}.
Utilization of the fast small matrix multiplication in practice is discussed in \cite{Benson,Ballard}.

Discrete or continuous optimization can be used as a method for canonical polyadic (CP) decomposition of the MM tensor. Either the decomposition is sought in the discrete domain, where the factor matrices may have elements only in the set $\{0,1,-1\}$, or the elements can be arbitrary, real or even complex-valued. One possibility to do the discrete optimization is to revert the problem to the problem of satisfiability of boolean equations \cite{SAT}, see, e.g., \cite{Courtois2,Heule}. In continuous optimization, the most remarkable results were obtained by Smirnov \cite{Smirnov}. More advanced decomposition methods have been proposed \cite{TDA,sensitivity}, and they are still being tested. Both discrete and continuous optimization might benefit from seeking symmetric decompositions \cite{Ballard2,Burichenko}.

Once one decomposition of the MM tensor is found, there is the whole group of De Groote transformations that lead to a whole class of equivalent solutions \cite{Groote}. An algorithm for determining wether two particular decompositions of the MM tensor are equivalent in the sense of the De Groote group has been proposed in \cite{Berger}. In Section IV of this paper, we extend the results of \cite{Berger} and suggest several characteristics that allow to identify equivalent
decompositions and, hopefully, to construct new decompositions.

The rest of the paper is organized as follows. The problem is formulated in Section II. A new formulation of the problem is proposed in Section 3.
The new formulation enables the simplification of the traditional ALS method of tensor decomposition, and reduces the complexity of computing the cost function.
The notion {\em signature} of decomposition is introduced in Section IV.
An ALS-based method is designed for finding a tensor decomposition with given signature. In Section V, another characteristic is proposed for decomposition of the MM tensor. It is called a rank signature.
How this signature helps to find decompositions with a lower number of nonzeros in the factor matrices is demonstrated. Section VI presents examples; the
most noteworthy is a novel decomposition of the MM tensor for the matrices
$3\times 3$ with $3\times 6$ which has rank 40.
Section VII concludes the paper.

{\bf Notation}. Boldface lowercase and uppercase letters will be used for vectors and matrices, respectively. Tensors are written in calligraphic letters, e.g., $\tT$. The corresponding matricizations along the mode $i$, $i=1,2,3$, will be denoted as $\bT_{(i)}$, respectively. Superscript $T$ denotes transpose, $\|\cdot\|_F$ represents the Frobenius norm of the argument (matrix or tensor), $\star$ is the elementwise (Hadamard) product,
$\odot$ is the Khatri-Rao product, $\otimes$ is the Kronecker product, $\bI$ represents the identity matrix, and
$\mbox{vec}(\cdot)$ is the operator of vectorization, which stacks all the elements of a matrix or a tensor in one column vector. Elements of a matrix $\bA$ are denoted $A_{ij}$, elements of a tensor $\tT$ are denoted $T_{ijk}$ or $T(i,j,k)$. Next, $\delta_{ij}$ is the Kronecker delta, and ${\bf 1}_N$ is the vector of ones of the size $N\times 1$.

\section{Tensor of Matrix Multiplications and its Decomposition}

Consider a bilinear mapping $\phi$ that represents the matrix product $\bZ=\bX\bY$, where
$\bX,\bY$, and $\bZ$ are real  or complex-valued matrices of the size $P\times Q$, $Q\times S$ and $P\times S$, respectively.
The mapping can be written symbolically as
\begin{equation}
\bz=\phi(\bx,\by)
\end{equation}
where $\bx=\mbox{vec}\, \bX^T$, $\by=\mbox{vec}\, \bY^T$, and $\bz=\mbox{vec}\, \bZ$, and $^T$ denotes the matrix transposition, and $\mbox{vec}$ means matrix vectorization.
The equation should hold for any matrices $\bX$ and $\bY$ and $\bZ=\bX\bY$. Indeed, the mapping $\phi$ is linear in both its arguments.

The mapping $\phi$ can be represented by a tensor $\tT_{PQS}$ of the size $PQ\times QS\times SP$ such that
\begin{equation}
\bz=\tT_{PQS}\times_1 \bx^T \times_2 \by^T
\end{equation}
where $\times_1$ and $\times_2$ denote multiplication along the first and the second dimension, which is written element-wise as
\begin{equation}
z_\gamma = \sum_{\alpha,\beta} T_{\alpha\beta\gamma} x_\alpha y_\beta~. \label{tri0}
\end{equation}
Matching (\ref{tri0}) with the definition relation of the matrix product
\begin{equation}
Z_{ps} = \sum_{q=1}^S X_{pq} Y_{qs}
\end{equation}
is achieved for $z_\gamma=Z_{ps}$, $x_\alpha=X_{pq}$, $y_\beta=Y_{qs}$, $\alpha=(p-1)Q+q$, $\beta=(s^\prime-1)S+s$, $\gamma=(s^\prime-1)P+p^\prime$, and
\begin{equation}
T_{\alpha\beta\gamma}=\delta_{pp^\prime}\delta_{qq^\prime}\delta_{ss^\prime}
\end{equation}
for all $p,p^\prime=1,\ldots,P$, $q,q^\prime=1,\ldots,Q$ and $s,s^\prime=1,\ldots,S$. 
The tensor has $P^2Q^2S^2$ elements but among them only $PQS$ nonzeros (ones). In other words, it is sparse.

In this paper, we study the canonical polyadic decomposition of the tensor in terms of factor matrices $\bA,\bB,\bC$,
where the rank of the decomposition $R$ is equal to the number of  columns in $\bA,\bB,\bC$. The matrices have the sizes
$PQ\times R$, $QS\times R$, and $PS\times R$, respectively.
Symbolically, we adopt the notation of \cite{Kolda}
\begin{equation}
\tT_{PQS}=[[\bA,\bB,\bC]]~.
\end{equation}
The CP decomposition means that elements of the tensor can be written as
\begin{equation}
T_{\alpha\beta\gamma}=\sum_{r=1}^R A_{\alpha r}B_{\beta r}C_{\gamma r}~.
\end{equation}
Given the factor matrices $\bA,\bB,\bC$, the matrix product $\bZ=\bX\bY$ can be written as
\begin{eqnarray}
z_\gamma = \sum_{\alpha,\beta} \sum_{r=1}^R A_{\alpha r}B_{\beta r}C_{\gamma r} x_\alpha y_\beta = \sum_{r=1}^R (\bA^T\bx)_r (\bB^T\by)_r C_{\gamma r}
\end{eqnarray}
or, in matrix form,
\begin{eqnarray}
\bz=\bC((\bA^T\bx)\star(\bB^T\by))\label{six}
\end{eqnarray}
where $\star$ denotes the elementwise (Hadamard) product.

Sometimes we need to represent the factor matrices in the form of order-3 tensors
and work with their slices. For example, $A(q,p,r)$ will denote the $((p-1)Q+q,r)$-th
element of $\bA$, $A(q,p,:)$ is the $[(p-1)Q+q]-$th row of $\bA$, and
$A(:,:,r)$ is the $r-$th column of $\bA$ reshaped in the matrix of the size $Q\times P$.
Similar notations will hold for the other factor matrices as well.

\section{New formulation for the CP decomposition}

Let us take the matrices $\bX$, $\bY$ at random, having i.i.d. ${\cal N}(0,1)$ distributed elements,
$\bz=\mbox{vec}(\bX\bY)$, $\bx=\mbox{vec}(\bX^T)$, $\by=\mbox{vec}(\bY^T)$, and compute the expectation
\begin{eqnarray}
\mbox{E}\|\bz-\bC((\bA^T\bx)\star(\bB^T\by))\|^2~.
\end{eqnarray}
The expectation is nonnegative and if it is zero, we have found a CP decomposition of the matrix multiplication tensor.
After a straightforward computation (see Appendix A) we
get\\
\noindent
{\bf Lemma 1}
\begin{eqnarray}
\mbox{E}\|\bz-\bC((\bA^T\bx)\star(\bB^T\by))\|^2=PQS-2\mbox{tr}(\bF^T\bC)+\mbox{tr}(\bC\bG\bC^T)\label{zde}
\end{eqnarray}
where $\bF=\bF(\bA,\bB)$ is a matrix of the size $PS\times R$ obtained by reshaping a tensor $\tF$ of the size $P\times S\times R$ defined through its elements
\begin{eqnarray}
F(p,s,:) = \sum_{q=1}^Q A(q,p,:)\star B(s,q,:)~.\label{rov2}
\end{eqnarray}
for $p=1,\ldots,P$, $q=1,\ldots,Q$, $s=1,\ldots,S$, and
\begin{eqnarray}
\bG &=& (\bA^T\bA)\star(\bB^T\bB)~.
\end{eqnarray}
\noindent
{\bf Proof}: See Appendix A.

As a consequence, we get\\
\noindent
{\bf Proposition 1}
For any matrices ${\bf A}$, $\bB$ and $\bC$ of the sizes $PQ\times R$, $QS\times R$, and $SP\times R$, respectively, it holds that
\begin{eqnarray}
PQS-2\mbox{tr}(\bF^T\bC)+\mbox{tr}(\bC\bG\bC^T)\geq 0~. \label{main}
\end{eqnarray}
An equality in (\ref{main}) happens if and only if $[[\bA,\bB,\bC]]=\tT_{PQS}$ is the CP decomposition of the
matrix multiplication tensor.\\
\noindent
{\bf Proof}\\
The inequality follows from the fact that the left-hand side is an expectation of the nonnegative random variable in (\ref{zde}).
If  $[[\bA,\bB,\bC]]=\tT_{PQS}$, then the random variable is zero with probability one. On the other hand, if the expectation
was strictly positive, the random variable almost certainly cannot be zero, and it would contradict (\ref{six}).
\hfill\rule{2mm}{2mm}

Interestingly enough, the criterion is numerically equivalent to the squared fitting error of the matrix
multiplication tensor.\\
\noindent {\bf Lemma 2}
\begin{eqnarray}
\phi(\bA,\bB,\bC)\stackrel{\triangle}{=} PQS-2\cdot\mbox{tr}(\bF^T\bC)+\mbox{tr}(\bC\bG\bC^T) = \|\tT_{PQS}-[[\bA,\bB,\bC]]\|_F^2 ~.\label{fce}
\end{eqnarray}
{\bf Proof}:
Note that
\begin{eqnarray}
\|\tT_{PQS}-[[\bA,\bB,\bC]]\|_F^2 &=& \|\tT_{PQS}\|_F^2 - 2\langle\tT_{PQS},[[\bA,\bB,\bC]]\rangle +  \|[[\bA,\bB,\bC]]\|_F^2
\end{eqnarray}
where $\langle\cdot,\cdot\rangle$ is a scalar product,
and
\begin{eqnarray}
\|\tT_{PQS}\|_F^2&=&PQS \label{jedna0} \\ \langle\tT_{PQS},[[\bA,\bB,\bC]]\rangle &=& \mbox{tr}(\bF^T\bC),\\  \|[[\bA,\bB,\bC]]\|_F^2 &=& \mbox{tr}(\bC\bG\bC^T)~.\label{treti}
\end{eqnarray}
See Appendix B For details.

Note that $\bF=\bF(\bA,\bB)$ is linear both in $\bA$, $\bB$, and $\bG=\bG(\bA,\bB)$ is quadratic. From (\ref{jedna0})-(\ref{treti}) it follows that if $[[\bA,\bB,\bC]]=\tT_{PQS}$
is the CP decomposition of the tensor, it holds that
\begin{eqnarray}
\mbox{tr}(\bF^T\bC) = \mbox{tr}(\bC\bG\bC^T) = PQS~.\label{PQS}
\end{eqnarray}
\\
\noindent {\bf Lemma 3}
The criterion in (\ref{fce}) is quadratic in $\bC$, and it can be minimized in closed form,
\begin{eqnarray}
\hat\bC=\bF\bG^{-1}~.\label{update}
\end{eqnarray}
Similarly, we can update $\bA$ given $\bB$ and $\bC$, as in the Alternating Least Squares (ALS) method for CP decomposition \cite{ALS,FALS}.
It is worth comparing the method with the standard ALS. The step of updating $\bC$ with fixed $\bA,\bB$ is
\begin{eqnarray}
\hat{\bC}=\bT_{(3)}(\bB\odot\bA)\bG^{-1}
\label{rov4}
\end{eqnarray}
where $\bT_{(3)}$ is the matricization of the tensor $\tT_{PQS}$ along its third mode, and $\odot$ is the Khatri-Rao product.
Since the target criterion is the same in both methods, we conclude that
\begin{eqnarray}
\bF=\bT_{(3)}(\bB\odot\bA)~.
\label{rov5}
\end{eqnarray}
This is because the tensor comprises of only $PQS$ ones, while the remaining elements are zeros.
The expression on the right-hand side of (\ref{rov5}) is known as the Matricized Tensor Times Khatri-Rao product (MTTKRP)
and it is usually the most computationally complex part of CP decomposition algorithms. In our case, computing $\bF$ according to (\ref{rov5}) requires $RP^2Q^2S^2$ flops, unless the tensor is stored in sparse representation. On the other hand, computing $\bF$ according to (\ref{rov2}) requires $RPQS$ flops only.
A similar decrease of complexity is found when computing the cost function
(\ref{fce}). The complexity of the traditional method in $RP^2Q^2S^2$ flops, but computing it in terms of $\bF$ and $\bG$ requires $RPQS+R^2(PQ+PS+SQ)$ flops only.

\section{The signature of the decomposition}

The CP decomposition of the MM tensor is, indeed, not unique. There is the trivial permutation and scale ambiguity, as in all CP decompositions,
but this is not all.
It holds that the set of triplets $\{(\bA,\bB,\bC)\}$ that represent CP decompositions of the MM tensor
is invariant with respect to the De Groote group of transformations which contains the following,
\begin{eqnarray}
(1) \qquad \bA^\prime \leftarrow (\bI_P\otimes\bX_A)\bA,&& \bB^\prime \leftarrow (\bX_A^{-T}\otimes\bI_T)\bB,\qquad \bC^\prime=\bC\label{trans}\\
(2) \qquad \bB^\prime \leftarrow (\bI_S\otimes\bX_B)\bB,&& \bC^\prime \leftarrow (\bX_B^{-T}\otimes\bI_P)\bC,\qquad \bA^\prime=\bA\label{trans2}\\
(3) \qquad \bC^\prime \leftarrow (\bI_S\otimes\bX_C)\bC,&& \bA^\prime \leftarrow (\bX_C^{-T}\otimes\bI_T)\bA,\qquad \bB^\prime=\bB,\label{trans3}
\end{eqnarray}
where $\bX_A,\bX_B,\bX_C$ are arbitrary invertible matrices of the sizes $Q\times Q$, $S\times S$, and $P\times P$, respectively, $\bI_P$, $\bI_Q$, and $\bI_S$ are the identity matrices of the indicated sizes, and
the decompositions are equivalent in the sense
\begin{eqnarray}
[[\bA^\prime,\bB^\prime,\bC^\prime]]&=&[[\bA,\bB,\bC]]~.
\end{eqnarray}
An intuitive explanation is that
the matrix multiplication $\bZ=\bX\bY$ is equivalent to the multiplications $\bZ=(\bX\bX_A)(\bX_A^{-1}\bY)$,
$\bX_B\bZ=(\bX_B\bX)\bY$, and $\bZ\bX_C=\bX(\bY\bX_C)$.
In other words, once we find one CP decomposition of the tensor, we obtain a whole variety of CP decompositions which can be parameterized
by elements of $\bX_A$, $\bX_B$, and $\bX_C$.

A natural question is whether all CP decompositions of the tensor are equivalent in the sense of the De Groote transformations. The answer is, indeed, negative,
except perhaps the simplest case with $P=Q=S=2$.\\
\noindent
{\bf Definition 1}\\
Let $[[\bA,\bB,\bC]]=\tT_{PQS}$ be a CP decomposition of the tensor.
We define the {\em signature} of the decomposition as
\begin{eqnarray}
\bs(\bA,\bB,\bC)={\bf 1}_{PS}^T(\bF\star\bC)~.
\end{eqnarray}
The signature is a vector of the length $R$ (the rank of the decomposition). Thanks to (\ref{PQS}), the sum of the signature elements is $PQS$, i.e., the number of ones in $\tT_{PQS}$.
The order of elements in the signature is not important: columns of the factor matrices $\bA,\bB,\bC$ can be re-ordered accordingly.
Without any loss of generality we can assume that
the signature elements are ordered in non-increasing fashion.

We claim\\ 
\noindent
{\bf Proposition 2}. The signature is invariant with respect to the De Groote group of transformations in the sense
\begin{eqnarray}
\bs(\bA^\prime,\bB^\prime,\bC^\prime)=\bs(\bA,\bB,\bC)~.
\end{eqnarray}
The proof is based on the fact that the $r-$th component of the signature
can be written as
\begin{equation}
s_r=\mbox{tr}\,\bH_r,\qquad r=1,\ldots,R
\end{equation}
where
\begin{equation}
\bH_r=\bC(:,:,r)\bB(:,:,r)\bA(:,:,r),\qquad r=1,\ldots,R~.\label{defH}
\end{equation}
In other words, $\bH_r$ is the product of the reshaped columns of the matrices $\bA,\bB,\bC$.
Another key observation is that the De Groote transformations convert these re-shaped columns as follows,
\begin{eqnarray}
\bA^\prime(:,:,r)&=&\bX_A\bA(:,:,r)\bX_C^{-1}\\
\bB^\prime(:,:,r)&=&\bX_B\bB(:,:,r)\bX_A^{-1}\\
\bC^\prime(:,:,r)&=&\bX_C\bC(:,:,r)\bX_B^{-1}~.
\end{eqnarray}
Then,
\begin{eqnarray}
\bH^\prime_r&=&\bC^\prime(:,:,r)\bB^\prime(:,:,r)^\prime\bA^\prime(:,:,r)=\bX_C\bH_r\bX_C^{-1}\label{similarity}
\end{eqnarray}
and
\begin{equation}
s_r^\prime=\mbox{tr}\,\bH^\prime_r=\mbox{tr}\,\bH_r =s_r \qquad r=1,\ldots,R~.
\end{equation}
We show later in the paper that not every two decompositions with the same signature are equivalent in the sense of De Groote group.

The signature has an interpretation revealed in the following theorems.\\
\noindent
{\bf Proposition 3}.
Let the tensor $\tT_{PQS}$ be written as a sum of $R$ rank-one components, $\tT=\sum_{r=1}^{R}\tT_r$
where $\tT_r=[[\bA_{:,r},\bB_{:,r},\bC_{:,r}]]$, then the $r-$th component of the signature is
$$
s_r=\langle\tT,\tT_r\rangle =\langle\tT,[[\bA_{:,r},\bB_{:,r},\bC_{:,r}]]\rangle~.
$$
\noindent
{\bf Proof}. It follows from Proof of Lemma 2 in Appendix B.

The Proposition can be understood in the following way. The tensor contains $PQS$ ones and the signature shows how many of these ones are covered by each of the  $R$ rank-one components.

\noindent
{\bf Proposition 4}.
Let the tensor $\tT_{PQS}$ have a CP decomposition $\tT_{PQS}=[[\bA,\bB,\bC]]$. Then, the signature of the decomposition is
also
\begin{eqnarray}
\bs(\bA,\bB,\bC)={\bf 1}_R^T [(\bA^T\bA)\star(\bB^T\bB)\star(\bC^T\bC)]={\bf 1}_R^T [\bG\star(\bC^T\bC)]~.
\end{eqnarray}
\noindent
{\bf Proof}. Thanks to Lemma 3 it holds that $\bC\bG=\bF$. Then, ${\bf 1}_R^T [\bG\star(\bC^T\bC)]={\bf 1}_{PT}^T [\bF\star\bC]=\bs(\bA,\bB,\bC)$.

\subsection{Examples}

There are some well known decompositions of the matrix multiplication tensors.
\begin{itemize}
\item Strassen's decomposition \cite{Strassen} of the tensor $\tT_{222}$ with rank 7 has the signature $(2,1,1,1,1,1,1)$.
\item Laderman's decomposition \cite{Laderman} of $\tT_{333}$ with rank 23 has the signature $(2,2,2,2,\underbrace{1,\ldots,1}_{19\times})$.
Similarly, Smirnov's decomposition of the tensor \cite{Smirnov} has the same signature.
\item Double Strassen decomposition of $\tT_{444}$ with rank 49 has the signature $(4,\underbrace{2,\ldots,2}_{12\times},\underbrace{1,\ldots,1}_{36\times})$.
\item Makarov's decomposition of $\tT_{555}$ \cite{Makarov} with rank 100 has the signature $(4,\underbrace{2,\ldots,2}_{22\times},\underbrace{1,\ldots,1}_{77\times})$.
\item Smirnov's decomposition of $\tT_{336}$ \cite{Smirnov} with rank 40 has the signature $\underbrace{\left(\frac{3}{2}\right.,\ldots,\frac{3}{2}}_{16\times},\underbrace{\frac{5}{4},\ldots,\left.\frac{5}{4}\right)}_{24\times}$
\end{itemize}
The first four decompositions have one aspect in common, namely that all factor matrices have, as elements, only $\{0,1,-1\}$.
Such decompositions are desired most. Signatures of such decompositions must be composed of integers.
In other words, if a decomposition with integer-valued factor matrices $\bA,\bB,\bC$ exists, the signature must be
integer-valued as well. It follows that no decomposition with elements in $\{0,1,-1\}$ which would be a De Groote-equivalent to the Smirnov's decomposition  exists.

When we conduct the numerical decomposition of the tensors, see, e.g. \cite{TDA}, we may obtain signatures that are not integer-valued. It seems that the signatures can be quite arbitrary.
Often they are close to integer-valued  vectors and then, it might be possible to find a decomposition with an integer-valued factor matrices. 

\subsection{Generalized Signature}

It was observed in \cite{Berger} that the matrices $\bH_r$ defined in (\ref{defH}) are similar to $\bH^\prime_r$
computed for $\bA^\prime,\bB^\prime,\bC^\prime$, see
(\ref{similarity}).
The similarity of the two matrices implies that the matrices have the same characteristic polynomial \cite{algebra}.
\begin{equation}
{\cal P}_r(t)=\mbox{det}(t\bI-\bH_r),\qquad t\in {\bf R}~.
\end{equation}
All coefficients of this polynomials are invariant to the De Groote transformation.
The signature element $s_r$ is the trace
of the matrix $\bH_r$, and  is just the first coefficient of the characteristic polynomial.
The other coefficients are invariant 
as well.
We call them the generalized signature: for each $r=1,\ldots,R$ it is a vector of coefficients of
${\cal P}_r(t)$.

It was also noted in \cite{Berger} that if an integer-valued decomposition exists, its generalized signature should be integer-valued as well. However, we observed that many times, the matrices $\bH_r$ have a low rank, and the higher elements of the generalized signature are zeros, and do not help to distinguish different decompositions.

\subsection{Decomposition with a given signature}

One might be interested in seeking a decomposition with a given signature. A typical situation where we need it is when we find 
a decomposition of the MM tensor which does not have an integer-valued signature, but the signature elements are close to some integers.
In that case it might be reasonable to seek a decomposition with an integer-valued signature, which is more interesting for practical reasons,
in a neighborhood of the given solution. Many times, this approach works, but not always. We tried, for example, to find a decomposition of the tensor
$\tT_{336}$ in a neighborhood of the known solution. It seems that no such solution exists.

Let the desired signature be $\bs_0$. The idea is to minimize, instead of the criterion (\ref{fce}), the
augmented criterion
\begin{eqnarray}
\varphi(\bA,\bB,\bC)=PQS-2\cdot\mbox{tr}(\bF^T\bC)+\mbox{tr}(\bC\bG\bC^T)+\|\bw^{1/2}\star(\bs_0-{\bf 1}^T(\bF\star\bC))\|^2\label{varphi}
\end{eqnarray}
where $\bw$ is a weight vector, so that each element of the signature has its own weight, and $\bw^{1/2}$ is the element-wise square root of $\bw$. As default, we can take $\bw$ as a vector of ones.

The criterion remains quadratic in $\bC$, and the minimization with respect to $\bC$ can be done in closed form again.

\noindent
{\bf Lemma 4}
\begin{eqnarray}
\mbox{vec}(\hat\bC)&=&\left[\bG\otimes\bI_{PS}+\mbox{diag}(\mbox{vec}(\bF))(\mbox{diag}(\bw)\otimes {\bf 1}_{PS,PS})\mbox{diag}(\mbox{vec}(\bF))\right]^{-1}\nonumber\\&&
[\mbox{vec}(\bF + \bF\mbox{diag}(\bs_0\star\bw))]~.\label{ALSs}
\end{eqnarray}
\noindent
{\bf Proof} See Appendix D.

Lemma 4 can be used in an alternating minimization method to find the CP decomposition with the given signature. An update of $\bA$ given the matrices $\bB$ and $\bC$, and an update of $\bB$ given $\bA$ and $\bC$ would be obtained from (\ref{ALSs}) by cyclic change $\bA\rightarrow\bB\rightarrow\bC\rightarrow\bA$. However, like in the ordinary ALS, the convergence might be slow and diverging solutions may be obtained, in which norms of the factor matrices go to infinity. More sophisticated algorithms exist \cite{TDA,KLM}.

\section{Rank Signatures}

In this section we introduce another kind of signature that is invariant with respect to
the De Groote group. We call it {\em rank signature} and it is computed for each
factor matrix separately. The definition is simple:
\begin{equation}
s^A_r=\mbox{rank}(\bA(:,:,r)),\qquad s^B_r=\mbox{rank}(\bB(:,:,r)),\qquad s^C_r=\mbox{rank}(\bC(:,:,r)),\qquad r=1,\ldots,R~.
\end{equation}
The rank signatures are always composed solely of integers.
To prove the invariance, note that the rank of a matrix remains constant when multiplying the matrix by another regular matrix from the left or from the right.

In some cases, the rank signatures may be equal to the ordinary signature, $s=s^A=s^B=s^C$. For  the Strassen algorithm
and double Strassen algorithm this condition holds, as it can be easily verified.  In general, however, the rank signatures are mutually different. We shall write the signatures in the form of a matrix with four rows: $s^A,s^B,s^C$, and the previous (vector) signature. For example, consider the following decomposition of tensor $\tT_{333}$. The origin of the decomposition will be explained in Section VI.

{\tiny
$$
\bA=\left(\begin{array}{cccccccccccccccccccccccc}
     0  &   1  &   1  &   0  &   0  &   0  &   0  &  -1  &   0  &   0  &   0   &  0 &   -1  &   0  &   0  &   0  &   0 &    0  &   0 &    0  &   0  &   1  & -1 \\
     0  &   0  &   0  &   1  &   0  &   0  &   1  &  -1  &   1  &   0  &   0   &  0  &   1  &   0  &  0   &  0  &   0  &   0   &  0   &  0  &   0   &  0   &  0 \\
     0  &   0  &   0  &   0  &   0  &   0  &   0  &   1  &   0   &  1  &   0   &  1 &    0 &    0  &   0  &   0  &   0  &   0  &   0  &   0 &    0  &   0  &   0 \\
     0  &   0  &   0  &   0  &   1  &   1  &   0  &  -1  &   1   &  0  &   1   &  0 &    0 &    0  &   0  &  -1  &   0 &   -1  &   0  &   1 &    0   &  0  &   0 \\
     1  &   0  &   0  &   0  &   0  &   0  &   1  &  -1  &   1  &   0  &   0   &  0 &    0  &   0  &   0  &   0  &   0 &    0  &   0  &   1 &    0   &  0  &   0 \\
     0  &   0  &   0  &   0  &   0  &   0  &   0  &   1  &  -1  &   0  &  -1   &  0 &    0  &   1  &   0  &   0  &   0 &    0  &   0  &  -1 &    0  &   0  &   0 \\
     0  &   0  &   0  &   0  &   0  &   0  &   0  &   0   &  0  &   0  &   0   &  0 &   -1  &   0  &   0  &   0 &    1 &   -1  &   0  &   0 &    0  &   1  &  -1 \\
     0  &   0  &   0  &   0  &   0  &   0  &   0  &   0   &  0  &   0  &   0   &  0 &    1  &   0  &   0 &    0  &   0&     1  &   1   &  0 &    1  &   0   &  1 \\
     0  &   1  &   0  &   0  &   0  &   0  &   0  &   0   &  0  &   1  &   0   &  1 &    0  &   1  &   1  &  -1  &  -1 &    0  &   0   &  0 &    0  &   0  &   0
\end{array}
\right)
$$
}
\begin{equation}
\bB=\bA(:,[1,2,10:23,3:9]),\qquad\bC=\bB(:,[1,2,10:23,3:9])\label{reordering}
\end{equation}
The above matlab notation means that $\bB$ and $\bC$ have the same columns as $\bA$, only in different permutation (order). The matrix signature of the decomposition is

{\tiny
$$
\bS=\left(\begin{array}{cccccccccccccccccccccccc}
1 &     2 &     1 &     1 &     1 &     1 &     1 &     1 &     2 &     1 &     1 &     1 &     1 &     1 &     1 &     2 &     1 &     2 &     1 &     1 &     1 &     1 &     2 \\
     1 &     2 &     1 &     1 &     1 &     1 &     1 &     1 &     2 &     1 &     2 &     1 &     1 &     1 &     1 &     2 &     1 &     1 &     1 &     1 &     1 &     1 &     2 \\
     1 &     2 &     1 &     2 &     1 &     1 &     1 &     1 &     2 &     1 &     1 &     1 &     1 &     1 &     1 &     2 &     1 &     1 &     1 &     1 &     1 &     1 &     2 \\
     1 &     2 &     1 &     1 &     1 &     1 &     1 &     1 &     2 &     1 &     1 &     1 &     1 &     1 &     1 &     2 &     1 &     1 &     1 &     1 &     1 &     1 &     2
\end{array}
\right)
$$
}
For comparison, the nonsymmetric decomposition of Smirnov of the same tensor in \cite{Smirnov}, has the same vector signature but a different rank signature. It has more rank-2 components in each factor matrix.

\subsection{Upper and lower bases}

In this subsection, we present a method of seeking a De-Groote-equivalent decomposition for an arbitrary decomposition of the MM tensor which might have more
zero entries in the factor matrices.

For each factor matrix in a CP decomposition of a MM tensor we suggest seeking two matrices, say $\bU_A$ and $\bL_A$ for
factor matrix $\bA$, $\bU_B$ and $\bL_B$ for
factor matrix $\bB$, and $\bU_C$ and $\bL_C$ for
factor matrix $\bC$ called upper and lower bases. The bases should have the minimum possible number of columns while fulfilling the property that each column of the factor matrix
is given as a linear combination of the products of the form $\bbu\odot \bbl$ where $\bbu$ is a column of $\bU_X$ and $\bbl$ is a column of $\bL_X$, X stands for A,B, and C, respectively. These bases are not bases in the ordinary sense in linear spaces, because they may contain more vectors than is the dimension of the space.

For example, assume that a column $\bA_r$ of $\bA$ has, after reshaping to $\bA(:,:,r)$, rank one. Then, $\bA(:,:,r)$
can be written as $\bA(:,:,r)=\bbl\bbu^T$, where $\bbu$ and $\bbl$
are column vectors of lengths $S$ and $P$, respectively. Then, $\bA_r=\bbu\odot \bbl$ and $\bbu$ or its scaled version should be found among columns of the upper basis $\bU_A$,
and $\bbl$ or its scaled version should exist among columns of the upper basis $\bL_A$.

 It can be easily computed that the symmetric decomposition of $\tT_{333}$ in the previous subsection has the upper and lower bases
 \begin{eqnarray}
\bL_A&=&\left(\begin{array}{cccccc}
 1 &  0 &     0 &   1 &    1 &  1 \\ 0 &  1 &     0 &   -1&    0 &  1 \\ 0&  0 &     1&   0 &    -1 &  -1
 \end{array}
\right) \label{base1}\\
\bU_A&=&\left(\begin{array}{cccccc}
 1 &  0 &     0 &   1 &    1 &  0 \\ 0 &  1 &     0 &   1 &    0 &  1 \\ 0 &  0 &     1 &   0 &    1 &  1
 \end{array}
\right)~. \label{base2}
\end{eqnarray}
 The other two factor matrices have the same upper and lower bases, because they differ only in the order of their columns.

The bases has the nice property that each of the De Groote transformations in (\ref{trans})-(\ref{trans3}) influences only two of them. For example, the transformation (\ref{trans}) influences
$\bU_A$ and $\bL_C$ so that
$\bU_A^\prime=\bX_A\bU_A$ and $\bL_C^\prime=\bX_A^{-1}\bL_C$.
The way that the columns of the factor matrices are formed from the bases (their linear coefficients) remains unchanged.
If the lower and upper bases contain an identity matrix as their submatrices, like the bases in (\ref{base1}) and  (\ref{base2}), there is probably no space for improvement. In general, however, the bases may contain no nulls. Then, a suitable choice of the transform matrices
$\bX_A$, $\bX_B$, $\bX_C$ would introduce identity matrices in the transformed bases, and consequently nulls in the factor matrices. The number of nulls in the factor matrices can be optimized by suitable selection of the base columns that are transformed to the identity.

\subsection{What else can be done}

Assume that we are given a sparse decomposition of the MM tensor in the sense that
the factor matrices contain many nulls, e.g., when obtained by the method proposed in the previous section. The decomposition might  not be satisfactory yet because it does not contain only integers or fractions of small integers.

In this case, it is advisable to apply a symbolic computation (Matlab, Mathematika, Maple, etc). The nonzero entries of the factor matrices are replaced by symbolic variables.
In order to reduce the number of the symbolic variables, we can normalize columns of two factor matrices so that each columns contains one "1", so that there is no longer any scale ambiguity. For easy reference, let us call the vector of the symbolic variables $p$. Then, we solve the system of equations
\begin{equation}
[[\bA(p),\bB(p),\bC(p)]]=\tT_{PQS}\label{Brent}
\end{equation}
for the unknown parameter $p$. The method consists of excluding one by one each unknown variable, writing it as a function of the remaining variables. If we are lucky, we can end up with a decomposition written as a function of just a few free variables. One such example is the following cyclically symmetric parametric decomposition of $\tT_{333}$.

{\tiny
$$
\bA(p)=\left(\begin{array}{cccccccccccccccccccccccc}
 1 &  0 &     0 &   0 &    0 &  0 &  0 &  0 &    0 &   0 &  0 &           1 &      1 &  0 &        0 &     0 &        0 &  0 &  0 &  0 &    0 &         0 &   0\\
 0 &  0 &     0 &   0 &    0 &  0 &  0 &  0 &    1 &   0 &  b &    -\frac{1}{df} &      d &  0 &        0 &     0 &  \frac{1}{df} &  1 &  0 &  0 &  b/f &         0 &   0\\
 0 &  0 &     a &  -a &    0 &  0 &  0 &  0 &    0 &   0 &  0 &          -b &      a &  0 &        1 &     0 &        0 &  0 &  1 &  0 &    0 &         0 &  -d\\
 0 &  0 &     0 &   0 &    0 &  1 &  0 &  0 &    f &   1 &  0 &        -\frac{1}{d} &    df &  0 &  \frac{1}{bd} &     0 &        0 &  0 &  0 &  0 &    0 &  -\frac{1}{bd} &   0\\
 0 &  1 &     0 &   0 &    0 &  0 &  0 &  0 &  df &   0 &  0 &           0 &  d^2f &  0 &        0 &     0 &        1 &  0 &  0 &  0 &    0 &      -1/b &   0\\
 0 &  0 &     0 &   0 &    0 &  0 &  0 &  1 &    0 &  -b &  0 &         b/d &      0 &  0 &     -1/d &     0 &        0 &  0 &  0 &  0 &    0 &         0 &   1\\
 0 &  0 &  -1/a &   0 &  1/a &  0 &  1 &  0 &    0 &   0 &  0 &         1/b &   -1/a &  0 &        0 &  -df &        0 &  0 &  0 &  0 &   -d &         0 &   0\\
 0 &  0 &     0 &   0 &    0 &  0 &  0 &  0 &    0 &   0 &  1 &  -\frac{1}{bdf} &      0 &  0 &        0 &     1 &        0 &  0 &  0 &  1 &  1/f &         0 &   0\\
 0 &  0 &    -1 &   1 &    1 &  0 &  0 &  0 &    0 &   0 &  0 &           0 &     -1 &  1 &        0 &     0 &        0 &  0 &  0 &  0 &    0 &         0 &   0
 \end{array}
\right)
$$
}
$$
\bB(p)=\bA(:,[1:5,12:23,6:11]),\qquad\bC(p)=\bB(:,[1:5,12:23,6:11])
$$
It can be easily verified that for any nonzero choice of $p=[a,b,d,f]$ we obtain a valid decomposition of the tensor.
Note that for any nonzero $p$, the decompositions have the same matrix signature, but they are not mutually De Groote-equivalent.
We can set $p$ to any combination of $\{1,-1\}$ to obtain a solution in $\{0,1,-1\}$.

\subsection{Symmetric CP decompositions}

The decomposition of MM tensors consists of optimizing a large number of parameters. The number of parameters can be reduced by utilizing special symmetries of the tensors. The tensors are symmetric and we can restrict our attention to symmetric decompositions only \cite{Ballard2,Burichenko}. The issue of symmetric decomposition exceeds the scope of this paper. Here we present only the cyclic symmetry mentioned in \cite{TDA}, because we need it in our examples.
The symmetry holds for multiplying two square matrices, $P=Q=S=N$. This tensor is invariant to cyclic permutation of its indices. In Matlab notation, it holds that
\begin{equation}
\tT_{NNN}=permute(\tT_{NNN},[2,3,1])=permute(\tT_{NNN},[3,1,2]) .
\end{equation}
In other words, if one decomposition of the tensor exists, $\tT_{NNN}=[[\bA,\bB,\bC]]$, then
\begin{equation}
\tT_{NNN}=[[\bA,\bB,\bC]]=[[\bB,\bC,\bA]]=[[\bC,\bA,\bB]]~.
\end{equation}
The decomposition is called symmetric if $\bA,\bB$ and $\bC$ differ only in the order of their columns.
In this case, $\bA,\bB$ and $\bC$ can be structured as
\begin{equation}
\bA=[\bA_0,\bA_1,\bB_1,\bC_1], \qquad \bB=[\bA_0,\bB_1,\bC_1,\bA_1], \qquad \bC=[\bA_0,\bC_1,\bA_1,\bB_1]\label{cyclic}
\end{equation}
where $\bA_1,\bB_1$, and $\bC_1$ are three matrices having the same number of columns. In this way, the number of independent parameters of the model is reduced to 1/3.

\section{Examples}

\subsection{Tensor $\tT_{333}$}

A few decompositions of the tensor have already been presented. The first one was obtained by discrete optimization using a SAT solver. In short, we solved the set of equations (\ref{Brent}) where $p$ contained all the elements of the factor matrix $\bA$ as independent unknown variables in the Galois field GF(2).
Recall that GF(2) contains
only two values, 0 and 1, and it holds that 1+1=0 in this field.
The other factor matrices were obtained by re-ordering the columns of $\bA$ according to
(\ref{reordering}). The system of equations (\ref{Brent}) for the unknown $p$
was converted into the SAT problem using the Bosphorus method and software
\cite{Bosphorus}, and solved via the solver Minisat 1.14 \cite{minisat}.
After cca 10 minutes of processing, we received a solution for $p$ in the Galois field.
The next step consisted of applying further symbolic computations described in
Section V.B. to assign suitable signs to the ones in $p$.

Finally, the columns  of $\bA$ were intentionally ordered so that the block had a nearly triangular form. The solution has 49 ones in each factor matrix.

The decomposition in Section V.B was obtained differently. The initial decomposition
of the tensor was obtained by the algorithm KLM with limited sensitivity, which was re-parametrized to respect the cyclical symmetry \cite{KLM}. Then, we used the technique
of lower and upper subspace from Section V.A to obtain a sparse solution, and finally the technique of symbolic computations from Section V.B. to find the decomposition with the minimum number of free parameters.

\subsection{Tensors $\tT_{444}$ and $\tT_{555}$}

A decomposition of $\tT_{444}$ with rank 49 can be found through a double application of the Strassen's algorithm. It has the cyclic symmetry (\ref{cyclic}) with $\bA_0=(1,0,0,0,0,1,0,0,0,0,1,0,0,0,0,1)^T$. Its signature is equal to the rank signature in all factor matrices, $s=s^A=s^B=s^C=(4,2,\ldots,2,1,\ldots,1)$.

The decomposition of $\tT_{555}$ with rank 100 was proposed by Makarov \cite{Makarov}. It depends on the decomposition of some smaller tensors, e.g., $\tT_{333}$. One possible realization
of the decomposition that we studied is posted on the Internet \cite{github}.
The decomposition is non-symmetric and has the signature $(4,2,\ldots,2,1,\ldots,1)$. Its rank signature is more complex, however.

\subsection{Tensor $\tT_{336}$}

In  \cite{Smirnov}, the author presents a CP decomposition of the tensor with rank 40 that contains ones, $1/8$'s and zeros in the factor matrices.
This solution can be converted through an appropriate De Groote transformation into another solution that contains only zeros, and plus or minus 1/2. It was converted through
the method proposed in \cite{TDA}, by numerically minimizing the L1 norm of the factor matrices via De Groote transformations.

Both the original and the transformed decompositions had the rank signatures
$$
s^A=(\underbrace{2,\ldots,2}_{16\times},\underbrace{1,\ldots,1}_{24\times}),\qquad s^B=s^C=(\underbrace{3,\ldots,3}_{40\times})~.
$$
Finally, we applied the method from Section VI.B of lower and upper bases of the factor matrices. Finding the bases of $\bA$ was easy, because there were rank-one columns.
Finding the bases for $\bB$  and $\bC$ was more tricky, because all columns of the matrices had a rank of 3.

What we did is that we sought the pairs $r,r^\prime$ such that the rank of the $6\times 6$ matrix $[\bB(:,:,r),\bB(:,:,r^\prime)]$ was 4. Then, we sought the vector in the columnspace of $\bB(:,:,r^\prime)$ that did not belong to the columnspace of $\bB(:,:,r)$, and added it to the upper base of $\bB$, if it was not already included. Thus, we had now selected 6 elements of the base to form the
transformation $\bX_B$. After applying it, the number of nonzero elements in the factor matrices $\bB$ and $\bC$ decreased significantly. Another reduction of the nonzero elements was obtained by a transformation $\bX_A$ from the lower basis of $\bA$.

The number of nonzero elements in the original decomposition and in the two novel decompositions are presented in the table below.

\begin{center}
\begin{tabular}{|c|c|c|c|}
\hline
& $\bA$& $\bB$& $\bC$\\ \hline
original & 192 & 384 & 384\\
novel A & 192 & 384 & 384\\
novel B & 144 & 192 & 312\\
\hline
\end{tabular}
\vspace*{2mm}

Table 1. Number of nonzero coefficients in decomposition of $\tT_{336}$.
\end{center}

Both novel decompositions were presented in Appendix E.

\section{Conclusions}

In the paper, a novel formulation of the decomposition of the matrix multiplication tensor was proposed.
The new formulation may help to accelerate numerical tensor decomposition procedures. Next, the signature of the decomposition was introduced. The signature can be used for  classification  and comparison of different decomposition results. We proposed a technique for seeking decompositions with integer-valued signatures, which are only able to provide decompositions in $\{0,1,-1\}$. Finally, we proposed the rank signature and a method for using it to seek sparse decompositions. A novel decomposition of the tensor
$\tT_{336}$ was derived that has a reduced number of nonzero entries in the factor matrices.
All decompositions mentioned in this paper were posted in electronic form in Matlab format \cite{github}.

\section*{Appendix A}

Proof of Lemma 1.
\begin{eqnarray}
\mbox{E}\|\bz\|^2&=& \mbox{tr}\,\mbox{E}(\bZ^T\bZ)=\mbox{tr}\,\mbox{E}(\bY^T\bX^T\bX\bY) = \mbox{tr}\,\mbox{E}(\bY^T\mbox{E}(\bX^T\bX)\bY)\nonumber\\
&=& P\, \mbox{tr}\,\mbox{E}(\bY^T\bY)=PQS~.
\end{eqnarray}
We used the fact that $\bX$ and $\bY$ are mutually independent, $\mbox{E}(\bX^T\bX)=P\bI_S$, and $\mbox{E}(\bY^T\bY)=T\bI_S$~.

Next,
\begin{eqnarray}
\mbox{E}\|\bC(\bA^T\bx\star\bB^T\by)\|^2&=& \mbox{tr}\{\bC\mbox{E}[(\bA^T\bx\star\bB^T\by)(\bA^T\bx\star\bB^T\by)^T]\bC^T\}\nonumber\\
&=& \mbox{tr}\{\bC\mbox{E}[(\bA^T\bx\bx^T\bA^T)\star(\bB^T\by\by^T\bB)]\bC^T\}\nonumber \\
&=& \mbox{tr}\{\bC[(\bA^T\bA)\star(\bB^T\bB)]\bC^T\}=\mbox{tr}(\bC\bG\bC^T)~.\label{42}
\end{eqnarray}
In the second equality of (\ref{42}) we used the fact that for any quartet of column vectors of the same size $\bv_1,\bv_2,\bv_3,\bv_4$ it holds that
$(\bv_1\star\bv_2)(\bv_3\star\bv_4)^T=(\bv_1\bv_3^T)\star(\bv_2\bv_4^T)=(\bv_1\bv_4^T)\star(\bv_2\bv_3^T)$.\\
Finally,
\begin{eqnarray}
&&\mbox{E}\{\bz^T\bC(\bA^T\bx\star\bB^T\by)\}\nonumber\\ &=& \mbox{E}\left\{\sum_{p,s} (XY)_{ps} \sum_r \bC(p,s,r)(\bA^T\bx\star\bB^T\by)_r\right\}\nonumber\\
&=& \mbox{E}\left\{\sum_{p,q,s} X_{pq}Y_{qs}\sum_r \bC(p,s,r)\left(\sum_{p^\prime,q^\prime}A(p^\prime,q^\prime,r)X_{p^\prime q^\prime}\right)\left(\sum_{q^{\prime\prime},s^\prime}B(q^{\prime\prime},s^\prime,r)Y_{q^{\prime\prime}s^\prime}\right)\right\}\nonumber\\
&=& \sum_{p,p^\prime,q,q^\prime,q^{\prime\prime},s,s^\prime}\delta_{p,p^\prime}\delta_{s,s^\prime}\delta_{q,q^{\prime\prime}}\delta_{tt^\prime} \sum_r C(p,s,r)A(p^\prime,s^\prime,r)B(q^{\prime\prime},s^\prime,r)
\nonumber\\
&=& \sum_{p,q,s,r} C(p,s,r) A(p,q,r) B(q,s,r)=\mbox{tr}(\bF^T\bC)~.
\end{eqnarray}

\section*{Appendix B}

Proof of Lemma 2.\\
Tensor $\tT_{PQS}$ contains $PQS$ ones, and the remaining elements are zero. Therefore $\|\tT_{PQS}\|_F^2=PQS$. Next,
\begin{eqnarray}
\|[[\bA,\bB,\bC]]\|_F^2&=&\|\bC(\bB\odot\bA)^T\|_F^2=\mbox{tr}[\bC(\bB\odot\bA)^T(\bB\odot\bA)\bC^T]\nonumber\\
&=&\mbox{tr}\{\bC[(\bA^T\bA)\star(\bB^T\bB)]\bC^T\}=\mbox{tr}(\bC\bG\bC^T)~.
\end{eqnarray}
Finally,
\begin{eqnarray}
\langle\tT_{PQS},[[\bA,\bB,\bC]]\rangle &=&\sum_{\alpha\beta\gamma} \tT_{\alpha\beta\gamma} ([[\bA,\bB,\bC]])_{\alpha\beta\gamma} \nonumber\\ &=&
\sum_{r=1}^R \sum_{\alpha\beta\gamma} \tT_{\alpha\beta\gamma} ([[\bA(:,r),\bB(:,r),\bC(:,r)]])_{\alpha\beta\gamma}\nonumber\\ &=&
\sum_{r=1}^R \sum_{p,p^\prime,q,q^\prime,s,s^\prime} \delta_{pp^\prime}\delta_{qq^\prime}\delta_{ss^\prime} A(p,q,r),B(q^\prime,s,r),C(p^\prime,s^\prime,r)\nonumber\\ &=&
\sum_{p,q,s,r} C(p,s,r) A(p,q,r) B(q,s,r)=\mbox{tr}(\bF^T\bC)~.
\end{eqnarray}

\section*{Appendix C}

Proof of Proposition 2.

The $r-$th column of the matrix $\bF$ is given as
\begin{eqnarray}
\bF_r = \mbox{vec} \,F(:,:,r) = \mbox{vec}\,[B(:,:,r) A(:,:,r)]^T
\end{eqnarray}
The $r-$th column of the matrix $\bA^\prime$ is given as
\begin{eqnarray}
\bA^\prime_r & =& (\bI\otimes \bX_A) \bA_r\nonumber\\
& =& (\bI\otimes \bX_A) \mbox{vec} A(:,:,r)\nonumber\\
& =& \mbox{vec} (\bX_A A(:,:,r))
\end{eqnarray}
Next,
\begin{eqnarray}
\bB^\prime_r & =& (\bX_A^{-T}\otimes \bI) \bB_r\nonumber\\
& =& (\bX_A^{-T}\otimes \bI) \mbox{vec} B(:,:,r)\nonumber\\
& =& \mbox{vec} (B(:,:,r)\bX_A^{-1})
\end{eqnarray}
Therefore,
\begin{eqnarray}
\bF^\prime_r &=& \mbox{vec} \,F^\prime(:,:,r) = \mbox{vec} \,[B^\prime(:,:,r) A^\prime(:,:,r)]^T\nonumber\\
& =& [B(:,:,r)\bX_A^{-1}\bX_A A(:,:,r)]^T = [B(:,:,r) A(:,:,r)]^T = \bF_r
\end{eqnarray}
The $r-$th element of the signature is given as
\begin{eqnarray}
s(r)&=&\sum_{p,s} F(p,s,r)C(p,s,r)\nonumber\\
&=&\sum_{p,t}\sum_s A(q,p,r)B(s,q,r)C(p,s,r)\nonumber\\
&=&\mbox{tr} [B(:,:,r)A(:,:,r)C(:,:,r)]~.
\end{eqnarray}
Obviously, the signature is invariant with respect to the cyclic change $A\rightarrow B\rightarrow C\rightarrow A$.
The statement of the Proposition follows.

\section*{Appendix D}

Proof of Lemma 3 and Lemma 4.\\
Put $\bff=\mbox{vec}\,\bF$ and $\bc=\mbox{vec}\,\bC$. The criterion in (\ref{update}) can be re-written as
\begin{eqnarray}
\phi(\bA,\bB,\bC)=PQS-2\bff^T\bc +\bc^T\mbox{vec}(\bC\bG)=PQS-2\bff^T\bc +\bc^T(\bG\otimes\bI_{PT})\bc~.\label{jedna}
\end{eqnarray}
Differentiation with respect to $\bc$ gives
\begin{eqnarray}
-2\bff +2(\bG\otimes\bI_{PT})\bc=0~.
\end{eqnarray}
The matrix $\bG$ is positive definite, therefore the minimum of $\phi(\bA,\bB,\bC)$ is achieved for
\begin{eqnarray}
\hat{\bc}=(\bG\otimes\bI_{PT})^{-1}\bff=(\bG^{-1}\otimes\bI_{PT})\mbox{vec}\,\bF=\mbox{vec}(\bF\bG^{-1})~.
\end{eqnarray}
The statement of Lemma 3 follows.\\
The criterion in (\ref{varphi}) has the additional term
\begin{eqnarray}
\|\bw^{1/2}\star[\bs_0-{\bf 1}_{PS}^T(\bF\star\bC)]\|^2&=&\|\bw^{1/2}\star\bs_0\|^2-2\cdot {\bf 1}_{PS}^T(\bF\star\bC)(\bw\star\bs_0)^T \nonumber\\&& +
\|\bw^{1/2}\star[{\bf 1}_{PS}^T(\bF\star\bC)]\|^2~.\label{dve}
\end{eqnarray}
Then,
\begin{eqnarray}
{\bf 1}_{PS}^T(\bF\star\bC)(\bw\star\bs_0)^T&=&(\bs_w\otimes{\bf 1}_{PS}^T)\mbox{vec}(\bF\star\bC)  \nonumber\\ &=&
(\bs_w\otimes{\bf 1}_{PS}^T)\mbox{diag}(\bff)\bc =\{\mbox{vec}[\bF\mbox{diag}(\bs_w)]\}^T\bc~.\label{tri}
\end{eqnarray}
where $\bs_w=\bs_0\star\bw$. Next,
\begin{eqnarray}
&&\|\bw^{1/2}\star[{\bf 1}_{PS}^T(\bF\star\bC)]\|^2=\|{\bf 1}_{PS}^T(\bF_w\star\bC)\|^2=\mbox{tr}[{\bf 1}_{PS}{\bf 1}_{PS}^T(\bF_w\star\bC)(\bF_w\star\bC)^T]\nonumber\\ &=&
\{\mbox{vec}[{\bf 1}_{PS,PS}(\bF_w\star\bC)]\}^T\mbox{vec}(\bF_w\star\bC)=\{\mbox{vec}(\bF_w\star\bC)\}^T[\bI_R\otimes{\bf 1}_{PS,PS}]\mbox{vec}(\bF_w\star\bC)\nonumber\\ &=& 
\bc^T\mbox{diag}(\bff)[\mbox{diag}(\bw)\otimes{\bf 1}_{PS,PS}]\mbox{diag}(\bff)\bc~.\label{ctyri}
\end{eqnarray}
where $\bF_w=\bF\mbox{diag}(\bw^{1/2})$. Combining (\ref{jedna}),(\ref{dve}),(\ref{tri}), and (\ref{ctyri}) we get, after some algebra, (\ref{ALSs}), as desired.

\section*{Appendix E}

Decomposition A of $\tT_{336}=[[\bA,\bB,\bC]]$.
{\tiny
\begin{eqnarray*}
\bA&=&\frac{1}{2}\left(\begin{array}{ccccccccccccccccccccc}
 1  &   1&    -1 &   -1   &  1 &   -1 &   -1  &   1  &  -1 &   -1  &   1  &   1 &   -1 &   -1 &   -1 &    1  &   0  &   1 &    0  &   0 \\
     0   &    0   &    0  &     0   &    0  &     0   &    0  &     0   &    0   &    0   &    0   &    0    &   0   &    0    &   0   &    0    &   0    &   1    &   0   &    1\\
     1   &   -1  &    -1  &     1   &    1   &    1   &   -1   &    1   &    1   &   -1   &    1   &    1   &    1   &    1    &   1    &  -1   &    0   &    0    &   0  &     1\\
     0   &    0  &     0   &    0   &    0   &    0   &    0  &     0   &    0    &   0  &     0   &    0   &    0   &    0   &    0   &    0   &   -1    &   0   &   -1  &     0\\
     1    &   1   &    1   &   -1   &    1   &    1   &    1  &     1   &    1   &    1  &    -1  &     1   &   -1  &     1   &    1   &    1   &   -1   &    0   &    1  &     1\\
     1   &   -1   &   -1   &   -1  &    -1   &   -1    &   1  &    -1   &   -1    &   1  &     1   &    1  &     1   &    1    &   1   &    1   &    0   &    0   &    0  &     1\\
    -1   &   -1   &    1   &    1   &   -1    &  -1    &  -1  &     1    &   1   &    1  &     1   &    1  &    -1   &   -1   &    1   &    1   &   -1  &    -1   &   -1  &     0\\
     1   &    1   &    1   &    1  &    -1    &   1    &   1  &     1    &  -1    &  -1  &    1   &   -1   &    1   &   -1    &   1   &    1   &   -1    &  -1   &    1   &    0\\
     0   &    0  &     0    &   0   &    0   &    0    &   0  &     0    &   0   &    0  &     0    &   0   &    0    &   0   &    0   &    0   &    0   &    0   &    0  &     0
\end{array}\right.\\&& \left.\begin{array}{ccccccccccccccccccccc}
   -1  &  -1  &   0   &  0  &  -1   &  0  &   1  &  -1  &   1  &   0   &  0   &  1  &   1   &  1  &   0  &   1  &   0  &   0   &  0 &   -1\\
     0  &   0  &   0  &   0  &   1  &  -1  &   0 &    0  &   0 &    0  &   0   &  0  &  -1  &   0  &   1 &    1  &   0  &   0  &  -1 &    0\\
     1 &   -1   &  0  &   0  &   0   &  1 &    1 &   -1 &   -1 &    0  &   0  &  -1  &   0  &  -1  &  -1  &   0  &   0 &    0  &  -1 &   -1\\
     1  &  -1  &   0  &   0  &   0  &   0 &    0 &    0  &   0  &   0  &   0  &   0 &    0  &   1  &   0  &   0  &  -1  &  -1  &   0 &    1\\
     0  &   0  &  -1  &  -1  &   0  &   1  &   0 &    0  &   0  &   1 &   -1  &   0  &   0  &   0  &   1  &   0   &  1  &  -1  &   1  &   0\\
    -1 &   -1  &   1  &  -1  &   0  &  -1  &   0  &   0  &   0  &  -1 &   -1   &  0  &   0  &  -1 &   -1   &  0  &   0  &   0   &  1  &   1\\
     0 &    0  &   0  &   0  &   1  &   0  &  -1  &  -1  &   1  &   0 &    0  &  -1 &    1 &    0  &   0  &   1  &   1  &   1  &   0  &   0\\
     0  &   0  &  -1  &  -1  &  -1   &  0  &   0  &   0  &   0  &  -1 &    1  &   0  &  -1  &   0  &   0   &  1   & -1   &  1   &  0  &   0\\
     0 &    0 &    1  &  -1  &   0   &  0   & -1  &  -1  &  -1  &   1 &    1  &   1 &    0   &  0  &   0  &   0   &  0   &  0   &  0  &   0
\end{array}\right)
\end{eqnarray*}
\begin{eqnarray*}
\bB&=&\frac{1}{2}\left(\begin{array}{ccccccccccccccccccccc}
     0 &     0 &    -1 &     0 &     0 &    -1 &    -1 &     0 &    -1 &    -1 &    -1 &     0 &     0 &     1 &     1 &     0 &    -1 &    -1 &    -1 &     1 \\
    -1 &     0 &     0 &    -1 &     0 &    -1 &     0 &     1 &     0 &     1 &     1 &     0 &    -1 &     0 &     1 &     0 &    -1 &     0 &    -1 &     0 \\
     0 &    -1 &     1 &     0 &    -1 &     0 &     1 &     0 &    -1 &     0 &     0 &    -1 &     0 &     1 &     0 &     1 &     1 &     0 &     1 &     0 \\
     0 &    -1 &     0 &     0 &     1 &     1 &     0 &     0 &     0 &    -1 &     1 &    -1 &     0 &     0 &     1 &    -1 &    -1 &     0 &     1 &     0 \\
     1 &    -1 &     0 &    -1 &     1 &     0 &     0 &    -1 &     0 &     0 &     0 &     1 &    -1 &     0 &     0 &     1 &     1 &    -1 &     1 &     1 \\
     1 &     0 &    -1 &    -1 &     0 &     0 &     1 &     1 &    -1 &     0 &     0 &     0 &     1 &    -1 &     0 &     0 &    -1 &     0 &     1 &     0 \\
     0 &     0 &    -1 &     0 &     0 &     1 &     1 &     0 &     1 &     1 &    -1 &     0 &     0 &     1 &     1 &     0 &    -1 &    -1 &     1 &     1 \\
     0 &     1 &     1 &     0 &    -1 &     0 &    -1 &     0 &     1 &     0 &     0 &     1 &     0 &     1 &     0 &     1 &     0 &     1 &     0 &     1 \\
     1 &     0 &     0 &    -1 &     0 &     1 &     0 &     1 &     0 &    -1 &     1 &     0 &     1 &     0 &     1 &     0 &     0 &    -1 &     0 &     1 \\
     1 &     0 &     1 &     1 &     0 &     0 &     1 &    -1 &    -1 &     0 &     0 &     0 &     1 &     1 &     0 &     0 &     0 &     1 &     0 &     1 \\
    -1 &     1 &     0 &    -1 &     1 &     0 &     0 &    -1 &     0 &     0 &     0 &    -1 &     1 &     0 &     0 &     1 &     1 &    -1 &    -1 &    -1 \\
     0 &    -1 &     0 &     0 &    -1 &     1 &     0 &     0 &     0 &    -1 &    -1 &    -1 &     0 &     0 &    -1 &     1 &     0 &     1 &     0 &    -1 \\
    -1 &    -1 &     0 &    -1 &    -1 &     0 &     0 &     1 &     0 &     0 &     0 &    -1 &    -1 &     0 &     0 &     1 &     0 &     0 &     0 &     0 \\
     0 &    -1 &    -1 &     0 &     1 &     0 &    -1 &     0 &    -1 &     0 &     0 &     1 &     0 &     1 &     0 &     1 &     0 &    -1 &     0 &     1 \\
     1 &     0 &     0 &    -1 &     0 &    -1 &     0 &    -1 &     0 &    -1 &    -1 &     0 &    -1 &     0 &     1 &     0 &     0 &    -1 &     0 &     1 \\
     0 &     1 &     0 &     0 &     1 &    -1 &     0 &     0 &     0 &    -1 &     1 &    -1 &     0 &     0 &    -1 &     1 &     1 &     0 &    -1 &     0 \\
     0 &     0 &     1 &     0 &     0 &    -1 &     1 &     0 &    -1 &     1 &     1 &     0 &     0 &     1 &     1 &     0 &     0 &     0 &     0 &     0 \\
     1 &     0 &    -1 &     1 &     0 &     0 &     1 &     1 &     1 &     0 &     0 &     0 &    -1 &     1 &     0 &     0 &    -1 &     0 &     1 &     0
\end{array}\right.\\&& \left.\begin{array}{ccccccccccccccccccccc}
   1 &     1 &     0 &     0 &     1 &    -1 &     0 &     0 &     0 &     0 &     0 &     0 &    -1 &     1 &    -1 &    -1 &    -1 &     1 &     1 &    -1 \\
     1 &    -1 &     0 &     0 &     0 &     0 &     0 &     0 &     0 &     0 &     0 &     0 &     0 &     1 &     0 &     0 &     1 &    -1 &     0 &     1 \\
     1 &    -1 &     0 &     0 &     0 &     0 &     0 &     0 &     0 &     0 &     0 &     0 &     0 &     1 &     0 &     0 &    -1 &     1 &     0 &     1 \\
    -1 &    -1 &     1 &     1 &     0 &     0 &     1 &     1 &     1 &     1 &     1 &     1 &     0 &     1 &     0 &     0 &    -1 &    -1 &     0 &    -1 \\
     1 &     1 &     0 &     0 &     1 &    -1 &     0 &     0 &     0 &     0 &     0 &     0 &    -1 &     1 &    -1 &    -1 &     1 &    -1 &     1 &    -1 \\
     1 &     1 &     1 &     1 &     0 &     0 &     1 &     1 &     1 &     1 &     1 &     1 &     0 &    -1 &     0 &     0 &    -1 &    -1 &     0 &     1 \\
    -1 &     1 &     0 &     0 &    -1 &     1 &     0 &     0 &     0 &     0 &     0 &     0 &     1 &     1 &    -1 &    -1 &     1 &     1 &    -1 &     1 \\
     0 &     0 &    -1 &    -1 &    -1 &     1 &    -1 &    -1 &    -1 &    -1 &    -1 &    -1 &    -1 &     0 &     1 &    -1 &     0 &     0 &     1 &     0 \\
     0 &     0 &     1 &     1 &     1 &     1 &     1 &     1 &     1 &     1 &     1 &     1 &     1 &     0 &     1 &     1 &     0 &     0 &     1 &     0 \\
     0 &     0 &     0 &     0 &     1 &    -1 &     0 &     0 &     0 &     0 &     0 &     0 &     1 &     0 &     1 &    -1 &     0 &     0 &    -1 &     0 \\
     1 &    -1 &     0 &     0 &    -1 &    -1 &     0 &     0 &     0 &     0 &     0 &     0 &     1 &    -1 &     1 &    -1 &    -1 &    -1 &     1 &    -1 \\
     0 &     0 &     0 &     0 &     1 &     1 &     0 &     0 &     0 &     0 &     0 &     0 &     1 &     0 &    -1 &    -1 &     0 &     0 &     1 &     0 \\
     0 &     0 &     1 &    -1 &     0 &     0 &    -1 &     1 &    -1 &    -1 &     1 &     1 &     0 &     0 &     0 &     0 &     0 &     0 &     0 &     0 \\
     0 &     0 &     1 &    -1 &     1 &    -1 &     1 &    -1 &    -1 &     1 &    -1 &     1 &    -1 &     0 &    -1 &    -1 &     0 &     0 &     1 &     0 \\
     0 &     0 &    -1 &     1 &     1 &    -1 &     1 &    -1 &    -1 &    -1 &     1 &     1 &    -1 &     0 &    -1 &    -1 &     0 &     0 &     1 &     0 \\
     1 &    -1 &    -1 &    -1 &     0 &     0 &     1 &     1 &    -1 &     1 &     1 &    -1 &     0 &    -1 &     0 &     0 &    -1 &    -1 &     0 &    -1 \\
     0 &     0 &    -1 &     1 &     0 &     0 &    -1 &     1 &    -1 &     1 &    -1 &     1 &     0 &     0 &     0 &     0 &     0 &     0 &     0 &     0 \\
    -1 &     1 &     1 &     1 &     0 &     0 &     1 &     1 &    -1 &    -1 &    -1 &    -1 &     0 &     1 &     0 &     0 &     1 &     1 &     0 &     1
\end{array}\right)
\end{eqnarray*}
\begin{eqnarray*}
\bC&=&\frac{1}{2}\left(\begin{array}{ccccccccccccccccccccc}
   -1 &     1 &     0 &    -1 &    -1 &     0 &     0 &     1 &     0 &     0 &     0 &    -1 &    -1 &     0 &     0 &    -1 &    -1 &    -1 &     1 &     1 \\
    -1 &     1 &     0 &     1 &     1 &     0 &     0 &    -1 &     0 &     0 &     0 &    -1 &    -1 &     0 &     0 &     1 &     1 &     1 &     1 &     1 \\
     0 &     0 &    -1 &     0 &     0 &     1 &     1 &     0 &    -1 &    -1 &    -1 &     0 &     0 &    -1 &     1 &     0 &     0 &     0 &     0 &     0 \\
    -1 &     0 &     0 &     1 &     0 &     1 &     0 &     1 &     0 &    -1 &     1 &     0 &     1 &     0 &    -1 &     0 &     0 &     1 &     0 &     1 \\
     0 &     1 &     1 &     0 &    -1 &     0 &    -1 &     0 &     1 &     0 &     0 &     1 &     0 &     1 &     0 &     1 &     1 &     0 &     1 &     0 \\
     1 &     0 &     0 &    -1 &     0 &     1 &     0 &     1 &     0 &     1 &     1 &     0 &     1 &     0 &     1 &     0 &     0 &    -1 &     0 &    -1 \\
     0 &    -1 &    -1 &     0 &    -1 &     0 &    -1 &     0 &     1 &     0 &     0 &    -1 &     0 &    -1 &     0 &     1 &     0 &    -1 &     0 &     1 \\
     1 &     0 &     0 &     1 &     0 &     1 &     0 &     1 &     0 &    -1 &    -1 &     0 &    -1 &     0 &     1 &     0 &    -1 &     0 &    -1 &     0 \\
     0 &     1 &     1 &     0 &     1 &     0 &    -1 &     0 &    -1 &     0 &     0 &    -1 &     0 &    -1 &     0 &     1 &     0 &     1 &     0 &     1 \\
     0 &    -1 &     0 &     0 &     1 &    -1 &     0 &     0 &     0 &     1 &     1 &    -1 &     0 &     0 &    -1 &    -1 &     0 &     1 &     0 &     1 \\
     1 &     0 &     1 &    -1 &     0 &     0 &     1 &    -1 &    -1 &     0 &     0 &     0 &    -1 &     1 &     0 &     0 &     1 &     0 &    -1 &     0 \\
     1 &     0 &     1 &     1 &     0 &     0 &     1 &    -1 &     1 &     0 &     0 &     0 &     1 &    -1 &     0 &     0 &     1 &     0 &    -1 &     0 \\
     0 &     0 &    -1 &     0 &     0 &    -1 &    -1 &     0 &    -1 &    -1 &     1 &     0 &     0 &     1 &     1 &     0 &    -1 &    -1 &     1 &    -1 \\
     0 &     0 &    -1 &     0 &     0 &     1 &     1 &     0 &     1 &     1 &     1 &     0 &     0 &     1 &     1 &     0 &    -1 &    -1 &    -1 &    -1 \\
    -1 &     1 &     0 &    -1 &    -1 &     0 &     0 &    -1 &     0 &     0 &     0 &     1 &     1 &     0 &     0 &     1 &     0 &     0 &     0 &     0 \\
     1 &     0 &     1 &     1 &     0 &     0 &    -1 &     1 &     1 &     0 &     0 &     0 &    -1 &     1 &     0 &     0 &     0 &     1 &     0 &    -1 \\
     0 &    -1 &     0 &     0 &    -1 &     1 &     0 &     0 &     0 &    -1 &     1 &    -1 &     0 &     0 &    -1 &     1 &     1 &     0 &    -1 &     0 \\
     0 &    -1 &     0 &     0 &     1 &     1 &     0 &     0 &     0 &     1 &    -1 &     1 &     0 &     0 &    -1 &     1 &     1 &     0 &    -1 &     0
\end{array}\right.\\&& \left.\begin{array}{ccccccccccccccccccccc}
    -1 &    -1 &     0 &     0 &    -1 &    -1 &     0 &     0 &     0 &     0 &     0 &     0 &    -1 &     1 &    -1 &    -1 &     1 &     1 &     1 &     1 \\
     1 &    -1 &     0 &     0 &    -1 &     1 &     0 &     0 &     0 &     0 &     0 &     0 &    -1 &     1 &    -1 &     1 &     1 &    -1 &    -1 &    -1 \\
     0 &     0 &     1 &     1 &     0 &     0 &     1 &    -1 &     1 &    -1 &     1 &     1 &     0 &     0 &     0 &     0 &     0 &     0 &     0 &     0 \\
     0 &     0 &     1 &    -1 &    -1 &    -1 &     1 &     1 &     1 &     1 &     1 &    -1 &     1 &     0 &     1 &    -1 &     0 &     0 &    -1 &     0 \\
     1 &     1 &     0 &     0 &     0 &     0 &     0 &     0 &     0 &     0 &     0 &     0 &     0 &     1 &     0 &     0 &    -1 &     1 &     0 &     1 \\
     0 &     0 &     1 &     1 &     1 &    -1 &    -1 &     1 &     1 &     1 &    -1 &     1 &     1 &     0 &     1 &    -1 &     0 &     0 &     1 &     0 \\
     0 &     0 &     1 &    -1 &     1 &    -1 &     1 &     1 &     1 &     1 &     1 &    -1 &    -1 &     0 &     1 &     1 &     0 &     0 &    -1 &     0 \\
     1 &     1 &     0 &     0 &     0 &     0 &     0 &     0 &     0 &     0 &     0 &     0 &     0 &     1 &     0 &     0 &     1 &    -1 &     0 &     1 \\
     0 &     0 &    -1 &    -1 &    -1 &     1 &    -1 &     1 &     1 &    -1 &     1 &     1 &    -1 &     0 &    -1 &     1 &     0 &     0 &    -1 &     0 \\
     0 &     0 &     0 &     0 &     1 &     1 &     0 &     0 &     0 &     0 &     0 &     0 &    -1 &     0 &     1 &    -1 &     0 &     0 &     1 &     0 \\
    -1 &     1 &    -1 &     1 &     0 &     0 &    -1 &    -1 &    -1 &    -1 &    -1 &     1 &     0 &     1 &     0 &     0 &     1 &     1 &     0 &    -1 \\
     1 &     1 &    -1 &     1 &     0 &     0 &    -1 &    -1 &     1 &     1 &     1 &    -1 &     0 &    -1 &     0 &     0 &    -1 &    -1 &     0 &    -1 \\
    -1 &    -1 &     0 &     0 &    -1 &     1 &     0 &     0 &     0 &     0 &     0 &     0 &    -1 &     1 &     1 &    -1 &     1 &     1 &    -1 &     1 \\
     1 &    -1 &     0 &     0 &     1 &    -1 &     0 &     0 &     0 &     0 &     0 &     0 &     1 &     1 &     1 &    -1 &    -1 &     1 &     1 &    -1 \\
     0 &     0 &    -1 &    -1 &     0 &     0 &     1 &    -1 &     1 &     1 &    -1 &     1 &     0 &     0 &     0 &     0 &     0 &     0 &     0 &     0 \\
     0 &     0 &     0 &     0 &     1 &    -1 &     0 &     0 &     0 &     0 &     0 &     0 &    -1 &     0 &    -1 &    -1 &     0 &     0 &    -1 &     0 \\
     1 &    -1 &     1 &    -1 &     0 &     0 &     1 &     1 &     1 &     1 &     1 &    -1 &     0 &    -1 &     0 &     0 &     1 &     1 &     0 &     1 \\
     1 &     1 &     1 &    -1 &     0 &     0 &    -1 &    -1 &     1 &    -1 &    -1 &    -1 &     0 &    -1 &     0 &     0 &    -1 &    -1 &     0 &    -1
\end{array}\right)
\end{eqnarray*}
}
Decomposition B of $\tT_{336}=[[\bA,\bB,\bC]]$.
{\tiny
\begin{eqnarray*}
\bA&=&\frac{1}{2}\left(\begin{array}{ccccccccccccccccccccc}
  0 &     1 &     0 &    -1 &     0 &    -1 &     0 &     0 &    -1 &     0 &     0 &     0 &    -1 &    -1 &    -1 &     1 &     0 &     0 &     0 &    -1 \\
     1 &     0 &    -1 &     0 &     1 &     0 &    -1 &     1 &     0 &    -1 &     1 &     1 &     0 &     0 &     0 &     0 &     0 &     1 &     0 &     1 \\
     0 &    -1 &     0 &     1 &     0 &     1 &     0 &     0 &     1 &     0 &     0 &     0 &     1 &     1 &     1 &    -1 &     0 &    -1 &     0 &     0 \\
    -1 &     0 &     0 &     1 &     0 &     0 &    -1 &     0 &     0 &    -1 &     0 &    -1 &     0 &    -1 &    -1 &    -1 &     0 &     0 &    -1 &    -1 \\
     1 &     0 &     0 &    -1 &     0 &     0 &     1 &     0 &     0 &     1 &     0 &     1 &     0 &     1 &     1 &     1 &    -1 &     0 &     0 &     1 \\
     0 &    -1 &    -1 &     0 &    -1 &    -1 &     0 &    -1 &    -1 &     0 &     1 &     0 &     1 &     0 &     0 &     0 &     1 &     0 &     0 &     0 \\
    -1 &    -1 &     0 &     0 &     0 &    -1 &    -1 &     0 &     1 &     1 &     0 &     1 &    -1 &     0 &     0 &     0 &     0 &     0 &    -1 &     0 \\
     0 &     0 &     1 &     1 &    -1 &     0 &     0 &     1 &     0 &     0 &     1 &     0 &     0 &    -1 &     1 &     1 &    -1 &    -1 &     0 &     0 \\
     0 &     0 &    -1 &    -1 &     1 &     0 &     0 &    -1 &     0 &     0 &    -1 &     0 &     0 &     1 &    -1 &    -1 &     1 &     1 &     0 &     0
\end{array}\right.\\&& \left.\begin{array}{ccccccccccccccccccccc}
   -1 &     0 &     0 &     0 &    -1 &     0 &     0 &     0 &     1 &     0 &     0 &     1 &     1 &     1 &     0 &     0 &     0 &     0 &     1 &     0 \\
     0 &    -1 &     0 &     0 &     0 &     0 &     1 &    -1 &     0 &     0 &     0 &     0 &     0 &     0 &     0 &     1 &     0 &     0 &    -1 &    -1 \\
     1 &     0 &     0 &     0 &     0 &     1 &     0 &     0 &    -1 &     0 &     0 &    -1 &     0 &    -1 &    -1 &    -1 &     0 &     0 &     0 &     0 \\
     1 &     0 &     0 &     1 &     0 &     0 &     0 &     0 &     0 &     0 &     1 &     0 &     0 &     1 &     0 &     0 &    -1 &     0 &    -1 &     0 \\
     0 &    -1 &     0 &    -1 &     0 &     0 &     0 &     0 &     0 &     0 &    -1 &     0 &     0 &     0 &     0 &     0 &     0 &    -1 &     1 &     1 \\
    -1 &     0 &     1 &     0 &     0 &    -1 &     0 &     0 &     0 &    -1 &     0 &     0 &     0 &    -1 &    -1 &     0 &     0 &     1 &     0 &     0 \\
     0 &     0 &     0 &     1 &     1 &     0 &     0 &     0 &     1 &     0 &    -1 &    -1 &     1 &     0 &     0 &     0 &     1 &     0 &     0 &     0 \\
     0 &     0 &     0 &    -1 &     0 &     0 &    -1 &    -1 &     0 &     0 &     1 &     0 &     0 &     0 &     0 &     1 &     0 &     1 &     0 &     0 \\
     0 &     0 &     1 &     0 &     0 &     0 &     0 &     0 &    -1 &     1 &     0 &     1 &     0 &     0 &     0 &    -1 &     0 &    -1 &     0 &     0
\end{array}\right)
\end{eqnarray*}
\begin{eqnarray*}
\bB&=&\left(\begin{array}{ccccccccccccccccccccc}
    0 &     0 &     0 &     0 &     0 &     0 &     0 &     0 &     0 &     0 &     0 &     0 &     0 &     0 &     0 &     0 &     0 &     0 &     1 &     0 \\
     0 &    -1 &     0 &     0 &     0 &    -1 &     0 &     0 &    -1 &     0 &     0 &     0 &    -1 &     0 &     0 &     0 &     0 &     0 &     0 &     0 \\
     1 &     0 &     0 &     0 &     0 &     0 &    -1 &     0 &     0 &    -1 &     0 &     1 &     0 &     0 &     0 &     0 &     0 &     0 &     0 &     1 \\
     1 &     0 &     0 &     0 &     0 &     0 &     1 &     0 &     0 &     1 &     0 &     1 &     0 &     0 &     0 &     0 &     0 &     0 &     1 &     1 \\
     0 &    -1 &     0 &     0 &     0 &     1 &     0 &     0 &     1 &     0 &     0 &     0 &    -1 &     0 &     0 &     0 &     0 &     0 &     1 &     0 \\
     0 &     0 &    -1 &    -1 &     1 &     0 &     0 &     1 &     0 &     0 &     1 &     0 &     0 &    -1 &     1 &    -1 &    -1 &    -1 &     0 &     0 \\
     1 &     0 &     0 &     0 &     0 &     0 &     1 &     0 &     0 &    -1 &     0 &    -1 &     0 &     0 &     0 &     0 &     0 &     0 &     1 &     0 \\
     0 &    -1 &     0 &    -1 &     0 &    -1 &     0 &     0 &    -1 &     0 &     0 &     0 &    -1 &     1 &     1 &     1 &     0 &    -1 &     0 &     1 \\
     0 &     0 &     0 &     0 &     0 &     0 &     0 &     0 &     0 &     0 &     0 &     0 &     0 &     0 &     0 &     0 &     0 &     0 &     0 &     0 \\
     1 &     0 &     0 &     0 &     0 &     0 &     1 &     0 &     0 &     1 &     0 &     1 &     0 &     0 &     0 &     0 &     0 &     0 &     1 &     0 \\
     0 &     0 &     1 &     0 &     1 &     0 &     0 &    -1 &     0 &     0 &     1 &     0 &     0 &     0 &     0 &     0 &     1 &     0 &     0 &     0 \\
     0 &     0 &    -1 &     0 &     1 &     0 &     0 &     1 &     0 &     0 &     1 &     0 &     0 &     0 &     0 &     0 &    -1 &     0 &     0 &     0 \\
     0 &     1 &     0 &     0 &     0 &    -1 &     0 &     0 &     1 &     0 &     0 &     0 &    -1 &     0 &     0 &     0 &     0 &     0 &     0 &     0 \\
     0 &    -1 &     0 &     0 &     0 &    -1 &     0 &     0 &    -1 &     0 &     0 &     0 &    -1 &     0 &     0 &     0 &     0 &     0 &     0 &     0 \\
     0 &     0 &     1 &     0 &    -1 &     0 &     0 &     1 &     0 &     0 &     1 &     0 &     0 &     0 &     0 &     0 &     0 &     1 &     0 &     0 \\
     1 &     0 &     0 &     1 &     0 &     0 &     1 &     0 &     0 &     1 &     0 &     1 &     0 &     1 &     1 &    -1 &    -1 &     0 &     1 &     1 \\
     0 &     0 &     0 &     0 &     0 &     0 &     0 &     0 &     0 &     0 &     0 &     0 &     0 &     0 &     0 &     0 &     0 &     0 &     0 &     0 \\
     0 &     0 &    -1 &     0 &     1 &     0 &     0 &     1 &     0 &     0 &     1 &     0 &     0 &     0 &     0 &     0 &     0 &    -1 &     0 &     0
\end{array}\right.\\&& \left.\begin{array}{ccccccccccccccccccccc}
    0 &     0 &     0 &     0 &    -1 &     0 &     0 &     0 &     0 &     0 &     0 &     0 &    -1 &     0 &     0 &     0 &    -1 &     0 &     0 &     0 \\
     1 &     0 &     0 &     0 &     1 &    -1 &     0 &     0 &     0 &     0 &     0 &     0 &    -1 &     1 &    -1 &     0 &     0 &     0 &     0 &     0 \\
     0 &     1 &     0 &     0 &     1 &     0 &     0 &     0 &     0 &     0 &     0 &     0 &    -1 &     0 &     0 &     0 &     0 &     0 &     1 &    -1 \\
     0 &     1 &     0 &     0 &     0 &     0 &     0 &     0 &     0 &     0 &     0 &     0 &     0 &     0 &     0 &     0 &     1 &     0 &    -1 &     1 \\
    -1 &     0 &     0 &     0 &     0 &     1 &     0 &     0 &     0 &     0 &     0 &     0 &     0 &     1 &    -1 &     0 &     1 &     0 &     0 &     0 \\
     0 &     0 &     1 &     1 &     0 &     0 &     1 &     1 &     1 &     1 &     1 &     1 &     0 &     0 &     0 &     1 &     0 &    -1 &     0 &     0 \\
     0 &     0 &     0 &     1 &     0 &     0 &     1 &     1 &     0 &     0 &     1 &     0 &     0 &     0 &     0 &     0 &    -1 &     0 &     0 &     0 \\
     1 &     0 &     0 &     0 &     1 &    -1 &     0 &     0 &    -1 &     0 &     0 &     1 &    -1 &     1 &    -1 &    -1 &     0 &     0 &     1 &     0 \\
     0 &     1 &     0 &     0 &     0 &     0 &    -1 &     1 &     0 &     0 &     0 &     0 &     0 &     0 &     0 &     0 &     0 &     0 &     0 &    -1 \\
     0 &     1 &     0 &     1 &     0 &     0 &     0 &     0 &     0 &     0 &    -1 &     0 &     0 &     0 &     0 &     0 &     1 &     0 &     0 &     1 \\
     0 &    -1 &    -1 &     0 &     0 &     0 &     0 &     0 &     0 &     1 &     0 &     0 &     0 &     0 &     0 &     0 &     0 &    -1 &     0 &    -1 \\
     0 &     0 &     1 &     0 &     0 &     0 &     1 &     1 &     0 &     1 &     0 &     0 &     0 &     0 &     0 &     0 &     0 &    -1 &     0 &     0 \\
     0 &     0 &    -1 &     0 &    -1 &     0 &     0 &     0 &    -1 &    -1 &     0 &    -1 &    -1 &     0 &     0 &     0 &     0 &     0 &     0 &     0 \\
     0 &     0 &     0 &     0 &     1 &    -1 &     0 &     0 &    -1 &     0 &     0 &     1 &    -1 &     0 &    -1 &     0 &     0 &     0 &     0 &     0 \\
     0 &     0 &     0 &     0 &     0 &     1 &    -1 &     1 &     0 &     0 &     0 &     0 &     0 &     0 &     1 &     1 &     0 &     0 &     0 &     0 \\
    -1 &     1 &     0 &     1 &     0 &     0 &     0 &     0 &     0 &     0 &    -1 &     0 &     0 &     1 &     0 &     0 &     1 &     1 &    -1 &     1 \\
     0 &     0 &    -1 &     0 &     0 &     1 &     0 &     0 &     0 &     1 &     0 &     0 &     0 &     0 &    -1 &     0 &     0 &     0 &     0 &     0 \\
     0 &     0 &     1 &     0 &     0 &     0 &     1 &     1 &     0 &     1 &     0 &     0 &     0 &     0 &     0 &     1 &     0 &     0 &     0 &     0
\end{array}\right)
\end{eqnarray*}
\begin{eqnarray*}
\bC&=&\frac{1}{2}\left(\begin{array}{ccccccccccccccccccccc}
   1 &    -1 &     0 &     0 &     0 &    -1 &    -1 &     0 &     1 &     1 &     0 &    -1 &    -1 &     0 &     0 &     0 &     0 &     0 &     0 &     0 \\
     1 &    -1 &     0 &     0 &     0 &     1 &     1 &     0 &    -1 &    -1 &     0 &    -1 &    -1 &     0 &     0 &     0 &     0 &     0 &    -2 &     0 \\
     0 &     0 &     1 &     1 &     1 &     0 &     0 &    -1 &     0 &     0 &    -1 &     0 &     0 &    -1 &    -1 &     1 &     1 &     1 &    -1 &     1 \\
    -1 &     0 &    -1 &     0 &    -1 &     0 &    -1 &     1 &     0 &    -1 &     1 &    -1 &     0 &     0 &     0 &     0 &    -1 &    -1 &     1 &     1 \\
     0 &     1 &     0 &     1 &     0 &     1 &     0 &     0 &     1 &     0 &     0 &     0 &    -1 &     1 &     1 &     1 &     0 &     0 &     0 &     0 \\
     0 &     1 &     0 &    -1 &     0 &     1 &     0 &     0 &    -1 &     0 &     0 &     0 &     1 &    -1 &     1 &     1 &     0 &     0 &     0 &     0 \\
     0 &     1 &     0 &    -1 &     0 &    -1 &     0 &     0 &    -1 &     0 &     0 &     0 &    -1 &     1 &     1 &    -1 &    -1 &    -1 &     1 &    -1 \\
    -1 &     0 &    -1 &     0 &     1 &     0 &     1 &    -1 &     0 &     1 &     1 &    -1 &     0 &     0 &     0 &     0 &     0 &     0 &     0 &     0 \\
    -1 &     0 &    -1 &     0 &    -1 &     0 &     1 &    -1 &     0 &    -1 &    -1 &     1 &     0 &     0 &     0 &     0 &     0 &     0 &     0 &     0 \\
     1 &     0 &     0 &     1 &     0 &     0 &    -1 &     0 &     0 &    -1 &     0 &     1 &     0 &     1 &     1 &     1 &     0 &     0 &     0 &    -2 \\
     0 &    -1 &    -1 &     0 &    -1 &     1 &     0 &     1 &     1 &     0 &     1 &     0 &     1 &     0 &     0 &     0 &    -1 &    -1 &    -1 &    -1 \\
    -1 &     0 &     0 &    -1 &     0 &     0 &    -1 &     0 &     0 &     1 &     0 &     1 &     0 &     1 &    -1 &     1 &     0 &     0 &     0 &     0 \\
     0 &    -1 &    -1 &     0 &     1 &    -1 &     0 &    -1 &    -1 &     0 &     1 &     0 &     1 &     0 &     0 &     0 &     0 &     0 &     0 &     0 \\
     1 &     0 &     0 &    -1 &     0 &     0 &     1 &     0 &     0 &     1 &     0 &     1 &     0 &     1 &     1 &    -1 &    -1 &    -1 &    -1 &    -1 \\
     0 &     1 &     1 &     0 &    -1 &    -1 &     0 &    -1 &     1 &     0 &     1 &     0 &     1 &     0 &     0 &     0 &     0 &     0 &     0 &     0 \\
     0 &     0 &     1 &     1 &     1 &     0 &     0 &     1 &     0 &     0 &     1 &     0 &     0 &     1 &    -1 &    -1 &     0 &     2 &     0 &     0 \\
     0 &     0 &     1 &    -1 &    -1 &     0 &     0 &    -1 &     0 &     0 &     1 &     0 &     0 &     1 &    -1 &     1 &     2 &     0 &     0 &     0 \\
     1 &    -1 &     0 &     0 &     0 &     1 &     1 &     0 &     1 &     1 &     0 &     1 &     1 &     0 &     0 &     0 &     1 &    -1 &    -1 &    -1
\end{array}\right.\\&& \left.\begin{array}{ccccccccccccccccccccc}
    0 &     0 &     0 &     0 &     2 &     0 &     0 &     0 &     0 &     0 &     0 &     0 &    -2 &     0 &     0 &     0 &     0 &     0 &     0 &     0 \\
     0 &     0 &     0 &     0 &     0 &     0 &     0 &     0 &     0 &     0 &     0 &     0 &     0 &     0 &     0 &     0 &     2 &     0 &     0 &     0 \\
     1 &     1 &    -1 &    -1 &    -1 &     1 &    -1 &    -1 &     1 &    -1 &     1 &    -1 &    -1 &    -1 &    -1 &     1 &    -1 &    -1 &    -1 &    -1 \\
    -1 &    -1 &     1 &    -1 &    -1 &    -1 &     1 &     1 &     1 &     1 &     1 &    -1 &    -1 &     1 &     1 &    -1 &     1 &     1 &    -1 &     1 \\
     2 &     0 &     0 &     0 &     0 &     0 &     0 &     0 &     0 &     0 &     0 &     0 &     0 &     2 &     0 &     0 &     0 &     0 &     0 &     0 \\
     0 &     0 &     0 &     0 &     0 &     0 &     0 &     0 &     2 &     0 &     0 &     2 &     0 &     0 &     0 &     0 &     0 &     0 &     0 &     0 \\
    -1 &    -1 &    -1 &     1 &    -1 &     1 &    -1 &    -1 &    -1 &    -1 &    -1 &     1 &    -1 &     1 &    -1 &    -1 &     1 &     1 &     1 &     1 \\
     0 &    -2 &     0 &     0 &     0 &     0 &     0 &     0 &     0 &     0 &     0 &     0 &     0 &     0 &     0 &     0 &     0 &     0 &     0 &    -2 \\
     0 &     0 &     0 &     0 &     0 &     0 &     2 &    -2 &     0 &     0 &     0 &     0 &     0 &     0 &     0 &     0 &     0 &     0 &     0 &     0 \\
     0 &     0 &     0 &     0 &     0 &     0 &     0 &     0 &     0 &     0 &     0 &     0 &     0 &     0 &     0 &     0 &     0 &     0 &    -2 &     0 \\
     1 &    -1 &     1 &    -1 &     1 &    -1 &     1 &     1 &     1 &     1 &     1 &    -1 &     1 &    -1 &     1 &    -1 &    -1 &     1 &     1 &     1 \\
     0 &     0 &     0 &    -2 &     0 &     0 &     0 &     0 &     0 &     0 &    -2 &     0 &     0 &     0 &     0 &     0 &     0 &     0 &     0 &     0 \\
     0 &     0 &     0 &     0 &     0 &     2 &     0 &     0 &     0 &     0 &     0 &     0 &     0 &     0 &     2 &     0 &     0 &     0 &     0 &     0 \\
    -1 &     1 &    -1 &     1 &     1 &    -1 &    -1 &    -1 &    -1 &    -1 &    -1 &     1 &     1 &     1 &     1 &    -1 &    -1 &     1 &     1 &    -1 \\
     0 &     0 &    -2 &     0 &     0 &     0 &     0 &     0 &     0 &     2 &     0 &     0 &     0 &     0 &     0 &     0 &     0 &     0 &     0 &     0 \\
     0 &     0 &     0 &     0 &     0 &     0 &     0 &     0 &     0 &     0 &     0 &     0 &     0 &     0 &     0 &    -2 &     0 &     0 &     0 &     0 \\
     0 &     0 &     0 &     0 &     0 &     0 &     0 &     0 &     0 &     0 &     0 &     0 &     0 &     0 &     0 &     0 &     0 &     2 &     0 &     0 \\
     1 &     1 &     1 &     1 &     1 &    -1 &    -1 &    -1 &     1 &     1 &    -1 &    -1 &     1 &    -1 &     1 &    -1 &    -1 &    -1 &     1 &    -1
\end{array}\right)
\end{eqnarray*}
}

\end{document}